\documentclass[conference]{IEEEtran}
\IEEEoverridecommandlockouts
\usepackage[utf8]{inputenc} 
\usepackage[T1]{fontenc}
\usepackage{epsfig,scalefnt,multirow,subfig}
\usepackage{url}
\usepackage{ifthen}
\usepackage{mathtools}
\usepackage{cite}
\usepackage{graphicx}
\usepackage{amssymb}
\usepackage{tabularx}
\usepackage{amsmath}
\usepackage{epstopdf}
\usepackage{algorithm}
\usepackage{algpseudocode}
\usepackage{algpascal}
\usepackage{cases}
\usepackage{lipsum}
\usepackage{stfloats}
\usepackage{float}
\usepackage{multicol}
\usepackage{xcolor}

  \def\cC{{\mathcal{C}}} 
\def\cE{{\mathcal{E}}}   
   
 \def\cN{{\mathcal{N}}}  
\def\cQ{{\mathcal{Q}}}

\def\argmin{\mathop{\mathrm{argmin}}}

\def\diag{\mathop{\mathrm{diag}}}

\def\trace{\mathop{\mathrm{tr}}}

\def\Re{\mathop{\mathrm{Re}}}
\def\Im{\mathop{\mathrm{Im}}}

\def\b0{{\pmb{0}}} 

  \def\bg{{\mathbf{g}}} \def\bh{{\mathbf{h}}}
   
 \def\bn{{\mathbf{n}}}  
\def\bq{{\mathbf{q}}} \def\br{{\mathbf{r}}} \def\bs{{\mathbf{s}}} 
  \def\bw{{\mathbf{w}}} 
\def\by{{\mathbf{y}}}   

\def\bA{{\mathbf{A}}}  \def\bC{{\mathbf{C}}} 
   \def\bH{{\mathbf{H}}}
\def\bI{{\mathbf{I}}}  \def\bK{{\mathbf{K}}} 
\def\bM{{\mathbf{M}}} \def\bN{{\mathbf{N}}}  
 \def\bR{{\mathbf{R}}}  
  \def\bW{{\mathbf{W}}} \def\bX{{\mathbf{X}}}
\def\bY{{\mathbf{Y}}}

\def\BibTeX{{\rm B\kern-.05em{\sc i\kern-.025em b}\kern-.08em
    T\kern-.1667em\lower.7ex\hbox{E}\kern-.125emX}}

\DeclarePairedDelimiter\norm{\lVert}{\rVert}
\begin{document}
\title{Channel Estimation for One-Bit Massive MIMO Systems Exploiting Spatio-Temporal Correlations} 


\author{%
  \IEEEauthorblockN{Hwanjin Kim and Junil Choi}
  \IEEEauthorblockA{Department of Electrical Engineering\\
                    Pohang University of Science and Technology\\
                    Pohang, Korea 37673\\
                    Email: \{jin0903, junil\}@postech.ac.kr}
 }


\maketitle

\begin{abstract}
 Massive multiple-input multiple-output (MIMO) can improve the overall system performance significantly. Massive MIMO systems, however, may require a large number of radio frequency (RF) chains that could cause high cost and power consumption issues. One of promising approaches to resolve these issues is using low-resolution analog-to-digital converters (ADCs) at base stations. Channel estimation becomes a difficult task by using low-resolution ADCs though. This paper addresses the channel estimation problem for massive MIMO systems using one-bit ADCs when the channels are spatially and temporally correlated. Based on the Bussgang decomposition, which reformulates a non-linear one-bit quantization to a statistically equivalent linear operator, the Kalman filter is used to estimate the spatially and temporally correlated channel by assuming the quantized noise follows a Gaussian distribution. Numerical results show that the proposed technique can improve the channel estimation quality significantly by properly exploiting the spatial and temporal correlations of channels.
\end{abstract}


\section{Introduction}\label{sec:introd}
Massive multiple-input multiple-output (MIMO) systems are a key approach for 5G wireless communication systems \cite{Marzetta06,Rusek13,Hoydis13,Larsson14}. Using large-scale antennas in base station (BS) can reduce inter-user interference by simple linear processing with {accurate} channel state information {at the BS and increase reliability by the channel hardening effect \cite{Marzetta06}.}

{Due to a large number of antennas, however, massive MIMO may suffer from high implementation cost and power consumption. It is possible to resolve these issues by using low-resolution analog-to-digital converters (ADCs) since the ADC power consumption exponentially increases with its resolution level \cite{Walden99}.} Recent work has revealed that massive MIMO using low-resolution ADCs can support multiple users transmitting high-order modulation symbols \cite{Wang15,Choi16,Mollen17,Mollen172}.

Successful symbol detection, however, requires accurate channel state information at the BS. Because low-resolution ADCs heavily quantize received signals, the channel estimation becomes an extremely difficult task. There has been some recent work to tackle this problem. A near maximum likelihood channel estimator based on convex optimization was proposed in \cite{Choi16}, and a joint channel and data estimator was developed in \cite{Wen16}. To reduce the channel estimation complexity, the generalized approximate message passing algorithm was exploited in \cite{Jmo17}, while the hybrid architectures was considered in \cite{Jmo171} for channel estimation. All the previous work, however, has not considered the temporal correlation, which is inherent in all communication channels. 

In this paper, we develop a novel channel estimator, which exploits both the spatial and temporal correlations of channels, for massive MIMO using one-bit ADCs. We first replace the non-linear one-bit quantizer to the linear operator by the Bussgang decomposition \cite{Bussgang52}. Then after approximating the statistically equivalent quantization noise as a Gaussian noise with the same mean and covariance matrix, we adopt the Kalman filter to exploit the temporal correlation and perform successive channel estimation \cite{Kalman}. The numerical results show that a normalized mean square error (NMSE) is decreased as the time slot increases. Moreover, as channels are more correlated, i.e., the spatial and temporal correlation coefficients are large, it is possible to estimate the channels more accurately. 

The rest of paper is organized as follows. In Section II, we describe a system model  using one-bit ADCs. In Section III, we explain the single-shot channel estimator with the Bussgang decomposition \cite{Li17}, then we propose the successive channel estimator with the Bussang decomposition and the Kalman filter. {In Section IV, we evaluate the achievable rate of the successive channel estimator. In Section V, we provide numerical results to evaluate the proposed channel estimator, and the conclusions follows in Section VI.}

\textbf{Notation:} Lower and upper boldface letters represent column vectors and matrices, respectively.  $\bA^{T}$, $\bA^{*}$, and $\bA^{H}$ denotes the transpose, conjugate, and conjugate transpose of the matrix $\bA$, respectively. $\mathbb{E}\{\cdot\}$ denotes the expectation and $\Re\{\cdot\}$, $\Im\{\cdot\}$ denotes the real part and imaginary part of the variable, respectively. $\boldsymbol{0}_m$ is used for the $m\times1$ all zero vector, and $\bI_m$ denotes the $m \times m$ identity matrix. $\otimes$ denotes the Kronecker product. $\diag(\cdot)$ returns the diagonal matrix. ${\mathbb{C}}^{m \times n}$ and ${\mathbb{R}}^{m \times n}$ represent the set of all $m \times n$ complex and real matrices, respectively. $|{\cdot}|$ denotes amplitude of the scalar and $\norm{\cdot}$ denotes the $\ell_2$-norm of the vector. $\cC \cN(m,\sigma^2)$ denotes the complex normal distribution with mean $m$ and variance $\sigma^2$.

\section{System Model}\label{sec:model}
\begin{figure}[t]
\centering
\includegraphics[width=0.45\textwidth]{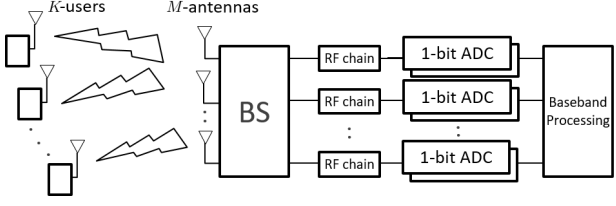} \caption{Massive MIMO systems with one-bit ADCs. Each RF chain has two one-bit ADCs for the real and imaginary parts, respectively, followed by the baseband processing}\label{fig:system}
\end{figure}

In Fig. \ref{fig:system}, we consider a MIMO system with $M$ BS antennas and $K$ single-antenna users. Each antenna is equipped with two one-bit ADCs for the real and imaginary parts, respectively. We assume the block-fading channel has a coherence time of $T$. At the $i$-th fading block, the received signal at the BS is given by 
\begin{align}
\by_i=\sqrt{\rho}\bH_i\bs_i+\bn_i \label {model}
\end{align}
where $\rho$  is the transmit SNR, $\bH_i=[\bh_{i,1},\bh_{i,2},...,\bh_{i,K}]$ is the $M\times K$ channel, $\bh_{i,k}$ is the channel between the $k$-th user and the BS in $i$-th fading block, $\bs_i$ is the transmitted signal,  and $\bn_i\sim \cC \cN(\boldsymbol{0}_{M},\bI_{M})$ is the noise. To model the spatially and temporally correlated channels, we assume $\bh_{i,k}$ follows the first-order Gauss-Markov process according to
\begin{align}
\bh_{0,k}&=\bR_k^{\frac{1}{2}}\bg_{0,k},\notag\\
\bh_{i,k}&=\eta_k\bh_{i-1,k}+\sqrt{1-\eta_k^2}\bR_k^{\frac{1}{2}}\bg_{i,k}, \quad i\geq1 \label{channel_model}
\end{align}
where $\bR_k={\mathbb{E}}\{\bh_{i,k}\bh_{i,k}^H\}$ is the spatial correlation matrix, $0\leq\eta_k\leq1$ is the temporal correlation coefficient, and $\bg_{i,k}\sim \cC \cN(\boldsymbol{0}_{M},\bI_{M})$ is the innovation process.

The quantized signal with the one-bit ADCs is
\begin{align}
\br_i=\mathcal{Q}(\by_i)=\mathcal{Q}( \sqrt{\rho}\bH_i\bs_i+\bn_i)
\end{align}
where $\mathcal{Q}(\cdot)$ is the one-bit quantization function as ${\mathcal{Q}}(\cdot)=\frac{1}{\sqrt{2}}(\mathrm{sign}({\Re}\{\cdot\})+j~\mathrm{sign}({\Im}\{\cdot\})).$
 
\section{Channel Estimation Using One-Bit ADCs}\label{sec:analysis}	
In this section, we first explain the conventional approach of signal-shot channel estimation without exploiting the temporal correlation of channels \cite{Li17}. Then, we propose a new successive channel estimation technique for massive MIMO using one-bit ADCs exploiting the temporal correlation. 

\subsection{Single-Shot Channel Estimator}\label{single}
In this subsection, we drop the time slot index $i$ since the signal-shot channel estimation does not exploit any temporal correlation. For the BS to estimate the channel, $K$ users simultaneously transmit their pilot sequences of $\tau$ symbols to the BS,
 \begin{align}
{\mathbf{Y}}=\sqrt{\rho}{\mathbf{H}}{\mathbf{\Phi}}^T+{\mathbf{N}}
\end{align}
where $\bY\in {\mathbb{C}}^{M\times\tau}$ is the receive signal, $\rho$ is the pilot transmit power, $\bH=[\bh_{1},\bh_{2},...,\bh_{K}]$ is the $M\times K$ channel, ${\mathbf{\Phi}} \in {\mathbb{C}}^{\tau \times K}$ is the pilot matrix and ${\mathbf{N}} \sim \mathcal{CN}({\boldsymbol{0}_M}, \bI_{M})$ is the Gaussian noise.
We assume that all pilot sequences are column-wise orthogonal, i.e. ${\mathbf{\Phi}}^T{\mathbf{\Phi}}^*=\tau \bI_K$. For simplicity, we vectorize the receive signal as
\begin{align}
{\mathrm{vec}}(\bY)=\by=\bar{\mathbf{\Phi}}\underline{\bh}+\underline{\bn} \label{y}
 \end{align}
 where $\bar{\mathbf{\Phi}}=(\mathbf{\Phi}\otimes\sqrt{\rho}\bI_M)$, $\underline{\bn} = \mathrm{vec}(\bN)$, and $\underline{\bh}=\mathrm{vec}(\bH)$. Note that the correlation matrix $\underline{\bR}={\mathbb{E}}\{\underline{\bh} \underline{\bh}^H\} $ is written as,
\begin{align}
\underline{\boldsymbol{R}}=\begin{bmatrix} 
&\boldsymbol{R}_1 &\cdots &0 &0\\
&\vdots &\boldsymbol{R}_2 &\cdots &0\\
&0 &\vdots &\ddots  &\vdots\\
&0 &0 &\cdots &\boldsymbol{R}_K \label{R_k}
\end{bmatrix} 
\end{align} 
assuming each user experiences an independent spatial correlation.
The quantized signal with one-bit ADCs is represented as
 \begin{align}
\br={\mathcal{Q}}(\by).\label{quantized}
 \end{align}

The previous channel estimators with one-bit ADCs in \cite{Choi16,Wen16,Jmo17} have high complexity due to the non-linear quantization of one-bit ADCs. The simple, yet effective, method is to use the Bussgang decomposition \cite{Bussgang52}. We review the channel estimator using the Bussgang decomposition proposed in \cite{Li17} because our channel estimator exploiting the temporal channel correlation relies on the techniques developed in \cite{Li17}. 

The Bussgang decomposition for one-bit quantizer is obtained by
 \begin{align}
\br={\mathcal{Q}}(\by)=\bA\by+\bq \label{Quantizer}
 \end{align}
where $\bA$ is the linear operator and $\bq$ is the statistically equivalent quantization noise \cite{Li17}. The linear operator $\bA$ is obtained by
\begin{align}
\bA&=\argmin _{\bA\in \mathbb{C}^{M\tau \times M\tau}}{\mathbb{E}}\left\{ {\|\br-\bA\by\|^2}\right\}\notag \\ 
   &=\argmin _{\bA\in \mathbb{C}^{M\tau\times M\tau}}\left(\bC_{\br}-\bC_{\br\by}\bA^H-\bA\bC_{\by\br}+\bA\bC_{\by}\bA^H\right) \notag\\ 
   &=\argmin_{\bA\in \mathbb{C}^{M\tau\times M\tau}}\Lambda(\bA)
 \end{align}
where $\bC_{\by\br}$ is the cross-covariance matrix between the receive signal $\by$ and the quantized signal $\br$, $\bC_{\br}$ is the auto-covariance matrix of $\br$, and $\bC_{\by}$ is the auto-covariance matrix of $\by$. 
The derivative of $\Lambda(\bA)$ with respect to $\bA^H$ is
\begin{align}
\frac{\partial\Lambda(\bA)}{\partial 
\bA^H}&=-\bC_{\by\br}^H+\bA\bC_{\by}=\boldsymbol{0}\notag \\
\bA&=\bC^H_{\by\br}\bC_{\by}^{-1},\label{linear_operator}
\end{align}
and $\bC_{\by\br}$ is given by \cite{Li17}
\begin{align}
\bC_{\by\br}=\sqrt{\frac{2}{\pi}}\bC_{\by}\diag(\bC_{\by})^{-\frac{1}{2}}.
\end{align}
The matrix $\bA$ is 
\begin{align}
\bA&=\bC^H_{\by\br}\bC_{\by}^{-1}=\left(\sqrt{\frac{2}{\pi}}\bC_{\by}\diag(\bC_{\by})^{-1/2}\right)^H\bC_{\by}^{-1}\notag \\
&=\sqrt{\frac{2}{\pi}}\diag(\bC_{\by})^{-\frac{1}{2}}\notag \\
&=\sqrt{\frac{2}{\pi}}\diag\left(\bar{\boldsymbol{\Phi}}\underline{\bR}\bar{\boldsymbol{\Phi}}^H+\bI_{M\tau}\right)^{-\frac{1}{2}}\notag\\
&\stackrel{(a)}{=}\sqrt{\frac{2}{\pi}}\diag(K\rho\underline{\bR}+\bI_{M\tau})^{-\frac{1}{2}}.\label{A}
\end{align}
 In (\ref{A}), (a) is because the diagonal terms of ${\boldsymbol{\bar{\Phi}}}{\boldsymbol{\bar{\Phi}}}^H$ is equal to $K \rho$. 
Using (\ref{y}) and (\ref{Quantizer}), $\br$ is represented as
\begin{align}
\br={\mathcal{Q}}(\by)=\tilde{{\boldsymbol{\Phi}}}\underline{\bh}+\tilde{\bn}
\end{align}
where $\tilde{\boldsymbol{\Phi}}=\bA\bar{\boldsymbol{\Phi}}\in{\mathbb{C}}^{M\tau\times{MK}}, \tilde{\bn}=\bA\bn+\bq\in{\mathbb{C}}^{M\tau\times 1}$.

Based on the Bussgang decomposition, we can construct the LMMSE estimator, which is referred as Bussgang LMMSE (BLMMSE) channel estimator \cite{Li17}:
\begin{align}
\underline{\hat{\bh}}^{\mathrm{BLM}}=\bC_{\underline{\bh}\br}\bC_{\br}^{-1}\br=\left(\bC_{\underline{\bh}}\tilde{\boldsymbol{\Phi}}^H+{\bC_{\underline{\bh}\bq}}\right)\bC_{\br}^{-1}\br\label{BLM}
\end{align}
where $\bC_{\underline{\bh}\br}$ is the cross-covariance matrix between $\underline{\bh}$ and $\br$.
The proof of (\ref{BLM}) is the same approach to obtain $\bA$ in (\ref{linear_operator}). 
$\bC_{\br}$ is given by the arcsin law \cite{arcsin}, which yields
\begin{align}
\bC_{\br}&=\frac{2}{\pi}\Big(\arcsin\Big(\Sigma_{\by}^{-1/2}\Re\{\bC_{\by}\}\Sigma_{\by}^{-1/2}\Big)\notag\\&+j\arcsin\Big(\Sigma_{\by}^{-1/2}\Im\{\bC_{\by}\}\Sigma_{\by}^{-1/2}\Big)\Big)\label{arcsinlaw}
\end{align}
where $\Sigma_{\by}=\diag(\bC_{\by})$.
Since $\bq$ is uncorrelated with $\underline{\bh}$ \cite{Li17}, the BLMMSE channel estimator of (\ref{BLM}) can be expressed as
\begin{align}
\underline{\hat{\bh}}^{\mathrm{BLM}}=\bC_{\underline{\bh}}\tilde{\boldsymbol{\Phi}}^H\bC_{\br}^{-1}\br. \label{BLM2}
\end{align}
We define the normalized mean squared error (NMSE) of the BLMMSE estimator as
\begin{align}
\mathrm{NMSE}_{\mathrm{BLM}}&=
\frac{1}{MK}{\mathbb{E}}\left\{ \left\|\hat{\underline{\bh}}^{\mathrm{BLM}}-\underline{\bh}\right\|_2^2\right\}\notag\\
&=\frac{1}{MK}{\mathrm{tr}}\left(\bC_{\underline{\bh}}-\bC_{\underline{\bh}}\tilde{\boldsymbol{\Phi}}^H\bC_{\br}^{-1}\tilde{\boldsymbol{\Phi}}\bC_{\underline{\bh}}^H\right). \label{BLMMSE_NMSE}
\end{align}

\subsection{Proposed Successive Channel Estimator}{\label{Successive}}
The proposed successive channel estimation technique that exploits the temporal channel correlation is based on the Bussgang decomposition, which is discussed in the previous subsection, and the Kalman filtering.\footnote{Note that we explicitly indicate the time slot index $i$ since the proposed channel estimator exploits the temporal correlation.} To develop the proposed channel estimator, we first reformulate the channel model in (\ref{channel_model}) using vectorized notations as
\alglanguage{pseudocode}
\begin{algorithm}
\caption{Channel Estimation Based on Kalman Filter }
\begin{algorithmic}[1]
    \State Initialization:
     \begin{align*}
     \hat{\underline{\bh}}_{0|-1}=\boldsymbol{0}_{MK},~
      {\bM}_{0|-1}=\underline{\bR}={\mathbb{E}}\left\{\underline{\bh}_0\underline{\bh}_0^H\right\}
     \end{align*}
     \State Prediction:
      \begin{align*}
     \underline{\hat{\bh}}_{i|i-1}=\underline{\boldsymbol{\eta}}\underline{\hat{\bh}}_{i-1|i-1}
      \end{align*}
      \State Minimum prediction MSE matrix ($MK\times MK$):
      \begin{align*}
     \bM_{i|i-1}=\underline{\boldsymbol{\eta}}\bM_{i-1|i-1}\underline{\boldsymbol{\eta}}^H+\underline{\boldsymbol{\zeta}}\underline{\bR}\underline{\boldsymbol{\zeta}}^H
      \end{align*}
      \State Kalman gain matrix ($MK\times M\tau$):
      \begin{align*}
     \bK_i=\bM_{i|i-1}\tilde{\boldsymbol{\Phi}}_i^H\left(\bC_{\check{\bn}_i}+\tilde{\boldsymbol{\Phi}}_i\bM_{i|i-1}\tilde{\boldsymbol{\Phi}}_i^H\right)^{-1}
      \end{align*}
      \State Correction:
      \begin{align*}
     \hat{\underline{\bh}}_{i|i}=\hat{\underline{\bh}}_{i|i-1}+\bK_i\left({\br_i}-\tilde{\boldsymbol{\Phi}}_i\hat{\underline{\bh}}_{i|i-1}\right)      \end{align*}
      \State Minimum MSE matrix ($MK\times MK$):
      \begin{align*}
     \bM_{i|i}=\left(\bI_{MK}-\bK_i\tilde{\boldsymbol{\Phi}}_i\right)\bM_{i|i-1}
      \end{align*}
\end{algorithmic}
\end{algorithm} 
\begin{align}
\underline{\bh}_0&=\underline{\bR}^{\frac{1}{2}}\underline{\bg}_0,\notag \\
\underline{\bh}_i&=\underline{\boldsymbol{\eta}}\underline{\bh}_{i-1}+\underline{\boldsymbol{\zeta}} \underline{\bR}^{\frac{1}{2}}\underline{\bg}_i, ~~~ i\geq1 \label{channel}
 \end{align}
 where $\underline{\bg}_i$ is the vectorized version of innovation process, which is represented as
\begin{align}
\underline{\bg}_i=\left[\bg_{i,1}^T,\bg_{i,2}^T,...,\bg_{i,K}^T\right]^T,~~~ i\geq0.
\end{align}
Note that the temporal correlation matrices $\underline{\boldsymbol{\eta}}$ and $\underline{\boldsymbol{\zeta}}$ in (\ref{channel}) are given by
\begin{align}
\underline{\boldsymbol{\eta}}&=\diag({\eta}_1,{\eta}_2,...,{\eta}_K) \otimes \bI_M,\notag \\
\underline{\boldsymbol{\zeta}}&=\diag({\zeta}_1,{\zeta}_2,...,{\zeta}_K) \otimes \bI_M
\end{align}
where $\eta_k$ is the temporal correlation coefficient of $k$-th user and $\zeta_k=\sqrt{1-\eta_k^2}$. 

We have the same step as in Section \ref{single} as the quantized signal is
 \begin{align}
\br_i={\mathcal{Q}}(\by_i).\label{quantized_i}
 \end{align}
The received signal with the Bussgang decomposition is
 \begin{align}
\br_i={\mathcal{Q}}(\by_i)=\bA_i\by_i+\bq_i \label{Quantizer_i}
 \end{align}
 where $\bA_i$ is the linear operator and $\bq_i$ is the statistically equivalent quantization noise. The receive signal can be represented as
\begin{align}
\br_i=\tilde{{\boldsymbol{\Phi}}}_i\underline{\bh}_i+\tilde{\bn}_i \label{receive_signal}
\end{align}
where $\tilde{\boldsymbol{\Phi}}_i=\bA_i\bar{\boldsymbol{\Phi}}_i\in{\mathbb{C}}^{M\tau\times {MK}}, \tilde{\bn}_i=\bA_i\bn_i+\bq_i\in{\mathbb{C}}^{M\tau\times 1}$.

The Kalman filter works when the noise is Gaussian distributed \cite{Kalman}; however, the effective noise $\tilde{\bn}_i$ in (\ref{receive_signal}) is not Gaussian. To over come this issue, we replace $\tilde{\bn}_i$ with $\check{\bn}_i$ that follows the Gaussian distribution with zero mean and the covariance matrix $\bC_{\tilde{\bn}_i}$, which is the covariance matrix of the effective noise $\tilde{\bn}_i$. We define a new receive signal $\check{\br}_i$,
\begin{align}
\check{\br}_i=\tilde{\boldsymbol{\Phi}}_i\underline{\bh}_i+\check{\bn}_i \label{check}
\end{align}
where $\check{\bn}_i\sim{\mathcal{CN}}(\boldsymbol{0}_{M\tau},\bC_{\tilde{\bn}_i})$.
The channel estimation based on the Kalman filter is summarized in Algorithm 1. Note that in Algorithm 1, $\underline{\bR}$ in Step 1 is defined in (\ref{R_k}), $\br_i$ in Step 5 is the quantized signal in (\ref{quantized_i}), not the approximated $\check{\br}_i$ in (\ref{check}).

\textbf{Remark: } The Gaussian approximation of the quantized noise may result in inaccurate tracking of channels. This effect becomes prominent especially in high SNR regime, which is shown in Fig. 4 in Section \ref{sec:numericalresult}. When SNR is high, the noise $\bn_i$ in (\ref{model}) becomes negligible and the effective noise $\tilde{\bn}_i$ in (\ref{receive_signal}) is dictated by the quantization process only, which would deviate from the Gaussian approximation further. If SNR is low, the effective noise $\tilde{\bn}_i$ becomes more like a Gaussian process, which makes the Kalman filtering perform well.

\section{Uplink Data Transmission}\label{sec:datatransmission}
In the data transmission stage, {the $K$ users} transmit data symbols to the BS. In the $i$-th time slot, the received signal at the BS can be represented as
\begin{align}
\br_{d,i}\notag&=\cQ(\by_{d,i})=\cQ(\sqrt{\rho_{d,i}}\bH_{i}\bs_{i}+\bn_{d,i})\\
&=\sqrt{\rho_{d,i}}\bA_{d,i}\bH_i\bs_i+\bA_{d,i}\bn_{d,i}+\bq_{d,i}\label{receivedata}
\end{align}
where $\bs_i$ is the transmit data symbol vector satisfying {${\mathbb{E}}\{|s_{i,k}|^2\}=1$, and} the subscript $d$ represents the data transmission. The linear operator for the Bussgang decomposition in (\ref{receivedata}) is given by
\begin{align}
\bA_{d,i}=\sqrt{\frac{2}{\pi}}\diag(\bC_{\by_{d,i}})^{-\frac{1}{2}}\stackrel{(a)}{\simeq}{\sqrt{\frac{2}{\pi}}\sqrt{\frac{1}{K\rho_{d,i}+1}}\bI_{M}}.\label{linearoperatordata}
\end{align}
{In (\ref{linearoperatordata}), we can approximate the matrix $\bA_{d,i}$ by (a) with a proper spatial correlation matrix satisfying $\sum_i\lambda_i=M$, where $\lambda_i$ is the $i$-th eigenvalue of $\bR_k$, as in \cite{Li17}.}
{The post-processed signal after receive combining is given by}
\begin{align}
\hat\bs_i\notag&=\bW_i^T\br_{d,i}\\
&=\sqrt{\rho_{d,i}}\bW_i^T\bA_{d,i}(\hat{\bH}_i\bs_i+\boldsymbol{\cE}_i\bs_i)+\bW_i^T\bA_{d,i}\bn_{d,i}+\bW_i^T\bq_{d,i}
\end{align}
where $\bW_i$ is the {receive} combiner, $\hat{\bH}_i =\text{unvec}({\hat{\underline{\bh}}_i})$ is the estimated channel matrix, and $\boldsymbol{\cE}_i=\bH_i-\hat{\bH}_i$ is the channel estimation {error matrix.}
{The $k$-th element of $\hat{\bs}_i$ can be expressed as}
\begin{align}
\hat{s}_{i,k}\notag&=\sqrt{\rho_{d,i}}\bw_{i,k}^T\bA_{d,i}\hat{\bh}_{i,k}s_{i,k}+\sqrt{\rho_{d,i}}\bw_{i,k}^T\sum_{j\neq k}^K\bA_{d,i}\hat{\bh}_{i,j} s_{i,j}\\
&+\sqrt{\rho_{d,i}}\bw_{i,k}^T\sum_{j=1}^K\bA_{d,i}\boldsymbol{\epsilon}_{i,j} s_{i,j}+\bw_{i,k}^T\bA_{d,i}\bn_{d,i}+\bw_{i,k}^T\bq_{d,i}
\end{align}
where $\bw_{i,k},\hat{\bh}_{i,k}$ and $\boldsymbol{\epsilon}_{i,k}$ are the {$k$-th} columns of $\bW_i,\hat{\bH}_i$, and $\boldsymbol{\cE}_i$, respectively.

\begin{figure*}[tbp]
	\begin{align}
	{R_k={\mathbb{E}}\left\{\log_2\left(1+\frac{\rho_{d,i}|\bw_{i,k}^T\bA_{d,i}\hat{\bh}_{i,k}|^2}{\rho_{d,i}\sum_{j \neq k}^K|\bw_{i,k}^T\bA_{d,i}\hat{\bh}_{i,j}|^2+\rho_{d,i}\sum_{j=1}^K|\bw_{i,k}^T\bA_{d,i}\boldsymbol{\epsilon}_{i,j}|^2+\norm{\bw_{i,k}^T\bA_{d,i}}^2+\bw_{i,k}^T\bC_{\bq_{d,i}}\bw_{i,k}^*}\right)\right\}\label{Rate}}
	\end{align}
\end{figure*}

{By treating the quantization noise $\bq_{d,i}$ as a Gaussian noise as in \cite{Diggavi01}, we can find a lower bound on the achievable rate of the $k$-th user in (\ref{Rate}). The covariance matrix of $\bq_{d,i}$ can be represented as
\begin{align}
{\bC_{\bq_{d,i}}}\notag&=\bC_{\br_{d,i}}-\bA_{d,i}\bC_{\by_{d,i}}\bA_{d,i}^H\\
               \notag&=\frac{2}{\pi}(\text{arcsin}(\bX)+j \text{arcsin}(\bY))-\frac{2}{\pi}(\bX+j\bY)\\
               &\stackrel{(a)}{\simeq}(1-2/\pi)\bI_M,\label{covariancematrixq}
\end{align}
{where we define}
\begin{align}
{\bX=\Sigma_{\by_{d,i}}^{-1/2}\Re\{\bC_{\by_{d,i}}\}\Sigma_{\by_{d,i}}^{-1/2},\bY=\Sigma_{\by_{d,i}}^{-1/2}\Im\{\bC_{\by_{d,i}}\}\Sigma_{\by_{d,i}}^{-1/2}}.
\end{align}
{In (\ref{covariancematrixq}), $\bC_{\br_{d,i}}$ can be obtained using the arcsin law in (\ref{arcsinlaw}), and (a) is from the low SNR approximation as in \cite{Li17}.}
The achievable sum-rate is given by}
\begin{align}
{R=\sum_{k=1}^K{R_k}.}
\end{align}
{Although any receive combiner is possible, we adopt the zero-forcing (ZF) combiner, which is given by} 
\begin{align}
{\bW_{i,ZF}^T=(\hat{\bH}_i^H\hat{\bH}_i)^{-1}\hat{\bH}_i^H,}
\end{align}
{for numerical studies.}

\section{Numerical Result}\label{sec:numericalresult}
In this section, we perform Monte-Carlo simulation to verify the proposed channel estimator. We use the NMSE as the performance metric,
\begin{align}
{\mathrm{NMSE}}=
\frac{1}{MK}{\mathbb{E}}\left\{ \left\|\hat{\underline{\bh}}-\underline{\bh}\right\|_2^2\right\}
\end{align}
where $\underline{\hat{\bh}}$ is the estimated channel.
We define the pilot sequence $\boldsymbol{\Phi}$ from the discrete Fourier transform (DFT) matrix and choose $K$ columns of $\tau\times\tau$ DFT matrix to get pilot sequence.
We adopt the exponential model for the spatial correlation matrix $\bR_k$:
 \begin{align}
 \bR_k=\begin{bmatrix} 
&1 &r_k &\cdots &r_k^{M-1}~\\
&r_k^* &1  &\cdots  &\vdots\\
&\vdots &\vdots &\ddots &\vdots\\
&r_k^{*(M-1)} &\cdots &\cdots &1~\\
\end{bmatrix}
\end{align}
where\footnote{We assume all users have the same $r$ (since it is a function of the BS antenna spacing) while each user experiences an independent phase $\theta_k$.} $r_k=re^{j\theta_k}$ ($0<r<1$,~$0<\theta_k<2\pi$). 
For the temporal correlation, we adopt Jakes' model, which given $\eta_k=J_0(2\pi f_Dt)$ where $J_0(\cdot)$ denotes the $0$-th order Bessel function, $f_D=vf_c/c$ is the Doppler frequency with the user speed $v$, the carrier frequency $f_c$, and the speed of light $c$, and $t$ is the channel instantiation interval \cite{Jakes}.


\begin{figure}[tbp]
\centering
\includegraphics[width=0.415\textwidth]{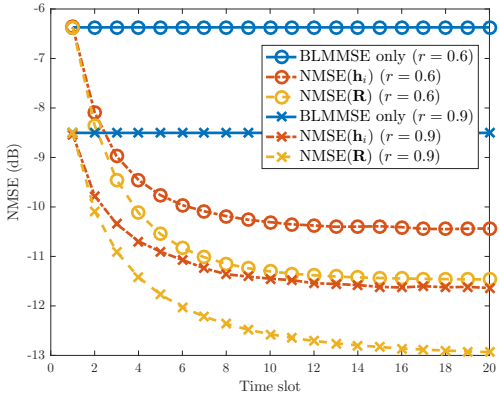}
\caption{{The NMSE} comparison according to time slot with different values of the spatial correlation coefficient $r$ when {$M=128$}, {$K=4$}, $\tau=20$, $\eta_k=0.9881$ and SNR = $-5$ dB.}\label{fig:NMSE1}
\end{figure}


\begin{figure}[tbp]
\centering
\includegraphics[width=0.415\textwidth]{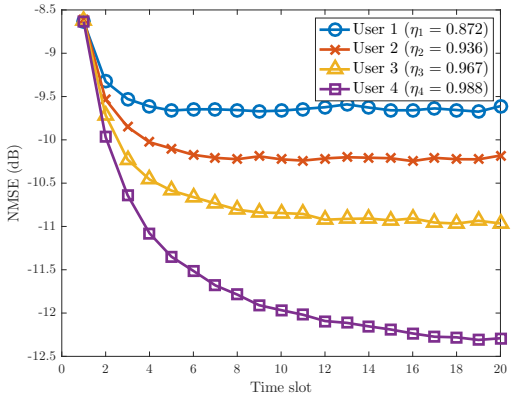}
\caption{{The NMSE} comparison according to time slot with user-dependent temporal correlations. {$M=128$}, $K=4$, $\tau=20$, $r=0.9$,  and SNR = $-5$ dB. The temporal correlation coefficients $\eta$ are $0.872, 0.936, 0.967$, and $0.988$ for each user.}\label{fig:NMSE2}
\end{figure}

In Fig. \ref{fig:NMSE1}, we compare the NMSE with respect to the time slot for $r=0.6$ or $r=0.9$ when SNR = $-5$ dB. We set the number of BS antennas 
{$M=128$}, the number of users {$K=4$}, and the number of training symbols $\tau=20$. The temporal correlation coefficient is $\eta_k=0.9881$ (which corresponds to $v=3\mathrm{km/h}$). We denote ${\mathrm{NMSE}}(\bh_i)$ as the NMSE performance of proposed technique at the $i$-th time slot.  Also, we denote ${\mathrm{NMSE}}(\bR)=\frac{1}{MK}\trace(\bM_{i|i})$ as the theoretical NMSE of Kalman filter with the Gaussian noise, not the true quantization noise where $\bM_{i|i}$ is the minimum MSE matrix defined in Step 6 of Algorithm 1. We also plot the BLMMSE only proposed in \cite{Li17}, which is the NMSE of the single-shot channel estimation explained in Section \ref{single}. In Fig. \ref{fig:NMSE1}, the NMSE of proposed channel estimator outperforms the BLMMSE only case with the time slot. The estimation performance of proposed technique becomes better when channels are more spatially correlated.

\begin{figure}[tbp]
\centering
\includegraphics[width=0.415\textwidth]{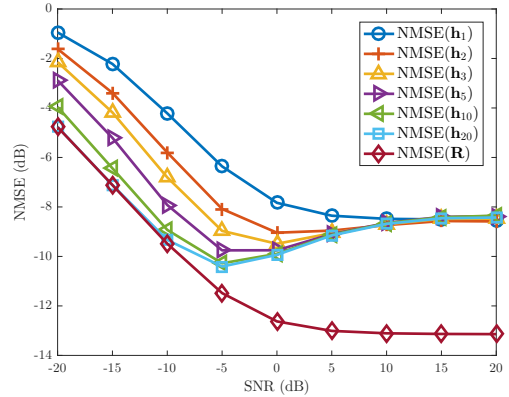}
\caption{{The NMSE} comparison according to SNR with different values of time slot $i$ when {$M=128$}, $K=4$, $\tau=20$, $\eta_k=0.9881$, and $r=0.6$.}\label{fig:NMSE3}
\end{figure}

In Fig. \ref{fig:NMSE2}, we compare the NMSE for each user when they experience user-dependent temporal fading. We set $r=0.9$ and the temporal correlation coefficient of each user as $\eta=0.872, 0.936, 0.967$, and $0.988$ (which correspond to $v_k=10\mathrm{km/h}, 7\mathrm{km/h}, 5\mathrm{km/h}$, and $3\mathrm{km/h}$). All other parameters are the same as in Fig. \ref{fig:NMSE1}. The figure shows that the proposed channel estimator works for the user-dependent temporal fading as well while the users experiencing more temporal correlation benefit more from the proposed estimator. 

In Fig. \ref{fig:NMSE3}, we compare the NMSE with respect to SNR with different values of the time slot index $i$. We set the temporal correlation coefficient $\eta_k=0.9881$ and the spatial correlation coefficient $r=0.6$. All other parameter are the same as in Fig. \ref{fig:NMSE1}. The proposed channel estimator has low NMSE with larger time slot index $i$. In high SNR regime, we verify that the one-bit ADCs quantizer has almost $5$ dB loss with respect to ${\mathrm{NMSE}}(\bR)$ where the loss comes from the Gaussian model mismatch as explained in \textbf{Remark} in Section \ref{Successive}. In low SNR regime, ${\mathrm{NMSE}}(\bh_i)$ is almost similar to ${\mathrm{NMSE}}(\bR)$ after 10 successive estimations. 
 
\begin{figure}[tbp]
	\centering
	\includegraphics[width=0.415\textwidth]{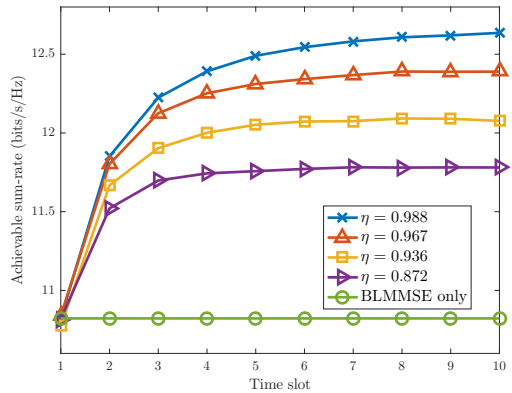}
	\caption{{The achievable rate} comparison according to time slot with different values of temporal correlations when {$M=64$}, {$K=4$}, $\tau=10$,  $r=0.6$, and SNR = 0 dB.}\label{fig:Sumrate_Time_SNR0}
\end{figure}
\begin{figure}[tbp]
	\centering
	\includegraphics[width=0.415\textwidth]{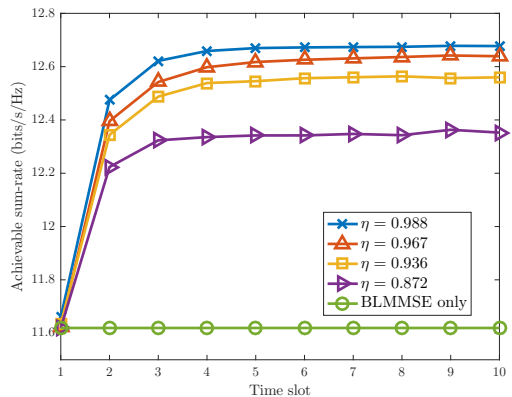}
	\caption{{The achievable rate} comparison according to time slot with different values of temporal correlations when {$M=64$}, $K=4$, $\tau=10$,  $r=0.6$, and SNR = 10 dB.}\label{fig:Sumrate_Time_SNR10}
\end{figure}

In Figs. \ref{fig:Sumrate_Time_SNR0} and \ref{fig:Sumrate_Time_SNR10}, we compare {the achievable sum-rate} according to the time slot with different temporal correlations when {$M=64$}, $K=4,\tau=10$ and SNR = 0 and 10 dB. We also plot the BLMMSE only case. {The achievable sum-rate} of proposed successive channel estimator outperforms the BLMMSE only case {as the time slot increases.} {The achievable sum-rate of proposed channel estimator increases with larger temporal correlation and SNR values.}

\section{Conclusion}\label{sec:conclusions}
In this paper, we proposed a channel estimation technique for spatially and temporally correlated channels in massive MIMO systems with one-bit ADCs. We exploited the Bussgang decomposition, which reformulates the non-linear function to statistically equivalent linear function, and the Kalman filter to estimate the channel by replacing the quantization noise to statistically equivalent Gaussian noise. The performance of proposed channel estimator has a substantial gain compared to the previous technique in \cite{Li17}, especially in low SNR regime. Also we verified that more accurate channel estimation is possible when channels are highly correlated in time and space. 

Possible future work includes deriving the fundamental performance limit of the proposed technique and implementing a dithering process to improve the performance in high SNR regimes \cite{Dither14}.


\section*{Acknowledgement}
This research was supported by the Institute for Information \& communications Technology Promotion (IITP) under grant funded by the MSIT of the Korea government (No.2018(2016-0-00123), Development of Integer-Forcing MIMO Transceivers for 5G \& Beyond Mobile Communication Systems) {and by the National Research Foundation (NRF) grant funded by the MSIT of the Korea government (2018R1A4A1025679).}

\IEEEtriggeratref{21}



\bibliographystyle{IEEEtran}
\bibliography{Globecom_ref}


\end{document}